\documentclass[showpacs,preprintnumbers,amsmath,amssymb]{revtex4}
\usepackage{graphicx} % Include figure files
\usepackage{dcolumn} % Align table columns on decimal point
\usepackage{bm} % bold math
\usepackage{epsfig}
\usepackage{amsfonts}

\begin{document}

\title{Comment on ``Proposed central limit behavior in deterministic dynamical systems''}

\author{Ugur Tirnakli}
\email{ugur.tirnakli@ege.edu.tr}
\affiliation{Department of Physics, Faculty of Science, Ege University, 35100 Izmir, Turkey}

\author{Constantino Tsallis}
\affiliation{Centro Brasileiro de Pesquisas F\'\i sicas\\
and National Institute of Science and Technology for Complex Systems,
Rua Dr. Xavier Sigaud 150, 22290-180
Rio de Janeiro, RJ, Brazil\\
and\\
Santa Fe Institute, 1399 Hyde Park Road, Santa Fe, NM 87501, USA}

\author{Christian Beck}
\affiliation{School of Mathematical Sciences, Queen Mary, University of London, Mile End Road,
London E1 4NS, UK }

\date{\today}

\begin{abstract}
In a recent Brief Report [Phys. Rev. E {\bf 79} (2009) 057201], Grassberger 
re-investigates probability densities of sums of iterates of
the logistic map near the critical point and claims that his simulation
results are inconsistent with previous results obtained by us
[U. Tirnakli {\it et al.}, Phys. Rev. E {\bf 75} (2007) 040106(R)
and Phys Rev. E {\bf 79} (2009) 056209]. In this comment we point out several errors
in Grassberger's paper.
We clarify that Grassberger's numerical simulations were mainly
performed in a parameter region that was explicitly excluded
in our 2009 paper and that his number of iterations is insufficient
for the region chosen.
We also show that, contrary to what is claimed by the author, 
(i)~L\'evy distributions are irrelevant for this problem,
and that (ii)~the probability distributions of
sums that focus on transients are unlikely to be universal.

\end{abstract}


\maketitle

In a recent Brief Report \cite{grass}, Grassberger claims that his numerical re-investigation
of the probability density of sums of iterates of the logistic map near to the critical
point of period doubling accumulation is inconsistent with our results previously
obtained in \cite{tibet1, tibet2}.
In \cite{tibet1} we provided for the first time numerical evidence for the possible relevance
of $q$-Gaussians for this problem, and in \cite{tibet2}
a more detailed investigation was performed with the main result
that
$q$-Gaussians are indeed a good approximation of the numerical data if
the parameter distance to the critical point and the number of iterations entering the sum
satisfy a scaling condition that was derived in \cite{tibet2}.
In \cite{grass} the author also claims that a) L\'evy distributions
could give an equally good fit to the data as $q$-Gaussians
b) L\'evy statistics might have a better theoretical basis
for this problem than $q$-statistics
c) new types of distributions that he obtains by
not neglecting transients could be universal.

In this note we point out 
that the paper \cite{grass}
is misleading since most of the
numerics performed in that paper operates in a parameter region
that we explicitly excluded by the scaling condition derived
in \cite{tibet2}. 
In the parameter region chosen by Grassberger, his statistics is
insufficient in the sense that much larger numbers of iterates
would be needed to observe $q$-Gaussian distributions.
Moreover we show that
claims a) and b) are incorrect and that there is no theoretical
or numerical basis for claim c).

Let us use the same notation as in \cite{grass}.
The object of study
are sums $Y$ of iterates $x_i$ of the logistic map
$f(x)=1-ax^2$ with parameter $a$ close to the critical point
$a_c=1.4011551890920506...$
of period doubling accumulation. The sum consists of $N$ iterates.
One starts from an ensemble of uniform initial conditions
and
the first $N_0$ iterates are ommitted:
\begin{equation}
Y= \sum_{i=N_0+1}^{N_0+N} x_i
\end{equation}
The question investigated in \cite{tibet1, tibet2, grass}
is what probability distribution is to be expected
for the random variable $Y$, since the
ordinary Central Limit Theorem (CLT) is not valid close to the critical
point due to strong correlations between the iterates.

Before discussing the results of \cite{grass}, we first point out
a few formal errors in \cite{grass}.
If $a$ is slightly
above the critical $a_c$ then it is well-known
that the attractor of the logistic
map consists of $n=2^k$ chaotic bands. In \cite{grass}
it is stated (1 line below caption of Fig. 3)
that $k \approx (a-a_c)^{1/\delta}$,
where $\delta$ is the Feigenbaum constant. This statement is obviously wrong,
the correct relation is
\begin{equation}
k= - \frac{\ln |a-a_c|}{\ln \delta}. \label{keq}
\end{equation}
There are a few further formal errors in \cite{grass}. In the caption of Fig. 4
in \cite{grass} it is said that the figure shows numerical data
for the probability density for various
values of $N$ and $n$. However, the actual data displayed
in Fig.~4
seem to correspond to tuples of the form
$(n,N)$, i.e., the order of $n$ and $N$ has been swapped. Another
error is the fact that the absolute
value is missing when the author refers to the $z$-logistic map $f_{a,z}(x)=a-|x|^z$
three lines after eq.(2).

Let us now come to the actual content of the paper.
In \cite{tibet2} we pointed out
that the problem is more complex
than the ordinary CLT, since two limits have to be performed
simultaneously: $a\to a_c$ and $N\to \infty$. Simultaneous
limits are standard knowledge in mathematics. In \cite{tibet2}
we provided arguments that in order to
obtain
$q$-Gaussian limit
distributions
the simultaneous limit $a \to a_c$
and $N\to \infty$ must be performed in such a way that the scaling relation
\begin{equation}
N \sim 4^{k} \label{3}
\end{equation}
holds.  Here $k$ is again given by eq.~(\ref{keq}),
and $\delta =4.6692011...$ is the Feigenbaum constant.
In the notation used by Grassberger
in \cite{grass} our scaling condition is equivalent to
\begin{equation}
N \sim n^2, \label{4}
\end{equation}
where again $n$ denotes the number of chaotic bands.
The abstract of \cite{grass} claims numerical inconsistency
but the paper is misleading since most of the simulations in \cite{grass}
ignore the above scaling condition (\ref{3}) or (\ref{4}) but
operate in a different parameter region (called `peaked region' in
\cite{tibet2}).
For example, for his results presented in
Fig.~2
of \cite{grass} the author
has chosen the distance $|a-a_c|$
from the critical point $a_c$ to be of the order $10^{-18}$,
basically fixed by his numerical precision.
For this value our scaling relation gives  $k =26.9$ and
hence $N \sim 4^k \sim 2^{54} \sim 1.5 \cdot 10^{16}$ is required to see a $q$-Gaussian.
On the contrary, the author performed his simulation in his Fig.~2 with
the iteration numbers $N=256, 2048, 16384$ and $131072$, which
are clearly insufficient to exhibit a $q$-Gaussian.
Being that close to the critical point, much higher values of iteration numbers
$N$ are needed to obtain sufficient statistics to properly confirm or disconfirm
our results presented in \cite{tibet2}. This
is precisely the reason why in \cite{tibet2}
we chose larger distances from $a_c$ for which the relevant $N$-values
are still reachable in a numerical experiment.

In Fig.~4 and 5 of \cite{grass}
the author investigates band splitting points
and mainly looks at cases in the parameter space
given by $n >> \sqrt{N}$ or $n << \sqrt{N}$, again ignoring our
scaling condition $n \sim \sqrt{N}$.
Singular behavior
of the density is observed
simply because the scaling condition (\ref{4}) is violated. 
Our present Fig.~1 shows how Fig.~4 of \cite{grass} would
have looked like had the scaling condition
been satisfied. In this case one gets smooth curves that are well-approximated
by $q$-Gaussians.
In this sense, for example the band splitting
point $4096\rightarrow2048$ investigated in \cite{grass}, which is the 
closest to the critical point, would yield the same $q$-Gaussian behavior
had the iteration number $N=2^{24}$ been used.

\begin{figure}
\includegraphics*[height=8cm]{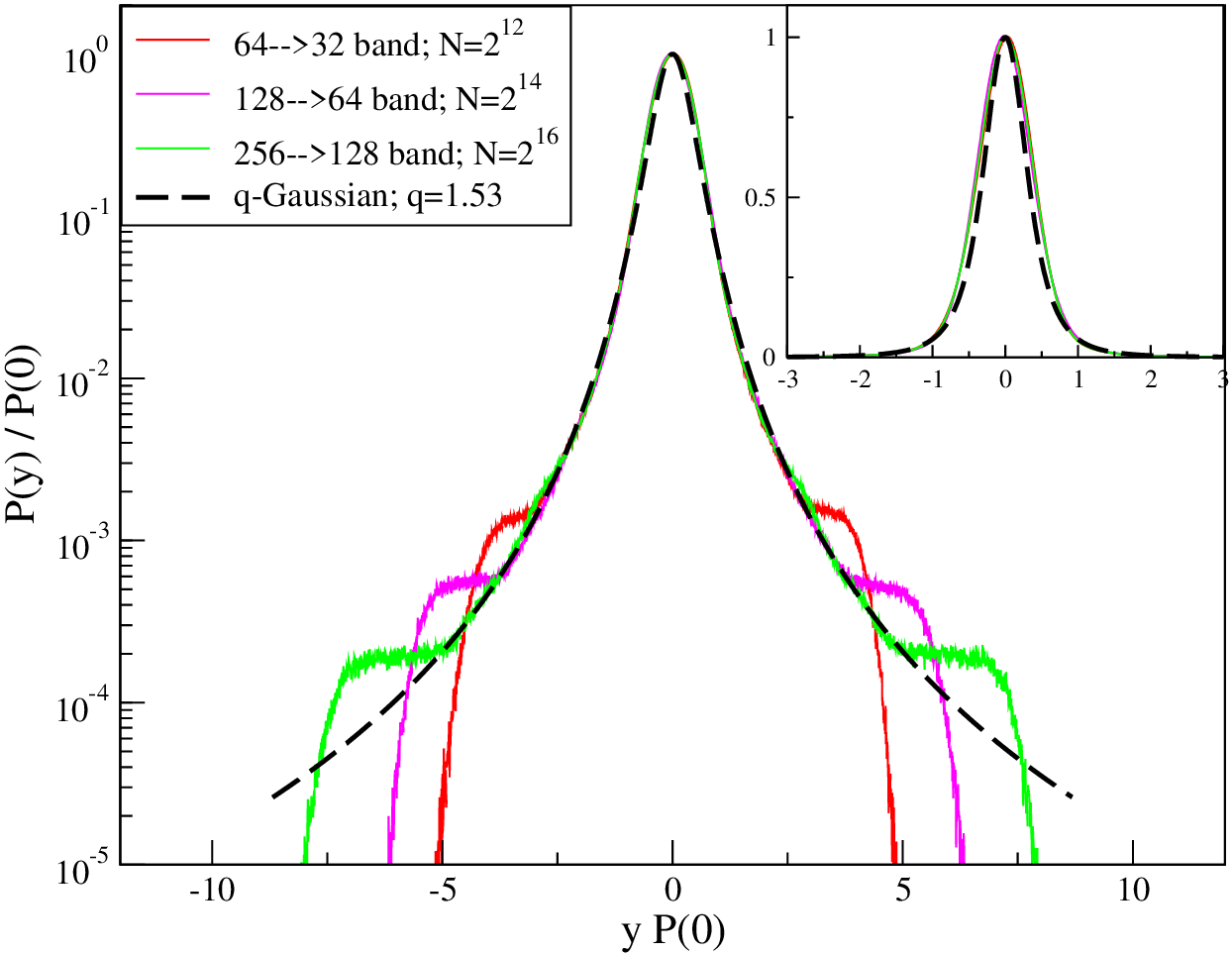}
\caption{Investigation of the
density of $Y$ for various band splitting points with number of
iterations $N$ satisfying the scaling condition
$N\sim n^2$.}
\end{figure}

Let us provide a few further comments.
In \cite{grass} some additional numerical experiments
were performed testing the effects of different lengths $N_0$
of omitted transients.
The author
emphasizes that in his opinion the condition $N_0>>N$ is relevant.
We have checked this claim
in the relevant parameter region
fixed by $N \sim n^2$.
As an example we have chosen one of the cases given in Fig.~2 
of \cite{tibet2} (namely, $a=1.401175$ and $N=16384$) and tested the effect
of discarding transients of length $N_0=2048,4096,8192,16384,65536$.
The result is shown in Fig.~2.
Apparently all curves fall
onto each other, no matter whether $N_0<N$ or $N_0>N$, and are well approximated by a
$q$-Gaussian. Hence the condition $N_0>>N$ advocated
in \cite{grass} seems to be irrelevant in the scaling region, as long as $N_0$ is sufficiently large
(in \cite{tibet1, tibet2} $N_0$ was typically chosen to be 4096).

\begin{figure}
\includegraphics*[height=8cm]{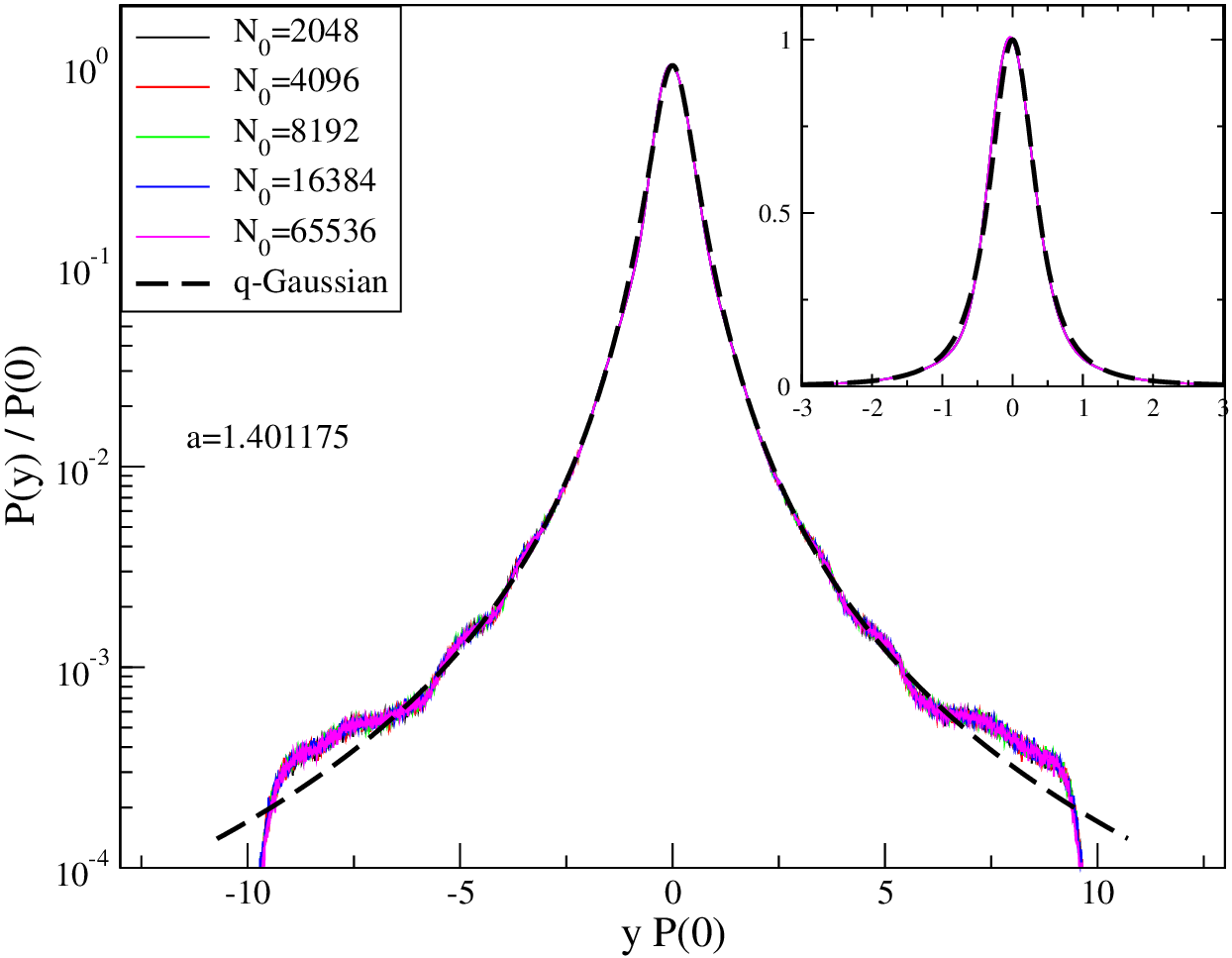}
\caption{Densities of $Y$
obtained for various lengths $N_0$ of omitted transients in the scaling
region $N \sim n^2$.}
\end{figure}

In \cite{grass} it is also stated that L\'evy distributions, possibly motivated by the
L\'evy-Gnedenko limit theorem, could give equally good fits.
To test this claim, we tried to fit our data by L\'evy distributions as well.
The result is shown in Fig.~3.
\begin{figure}
\includegraphics*[height=8cm]{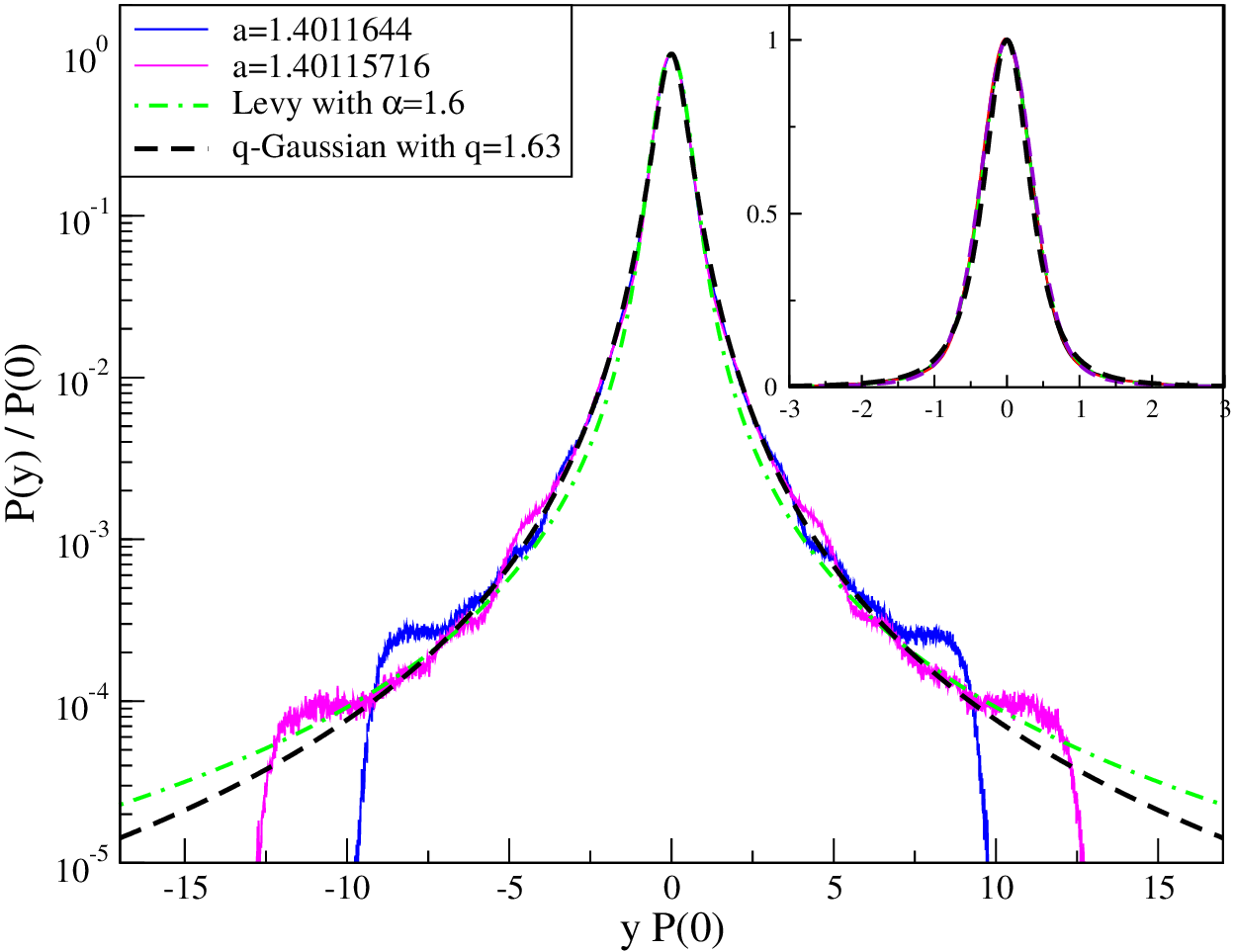}
\includegraphics*[height=8cm]{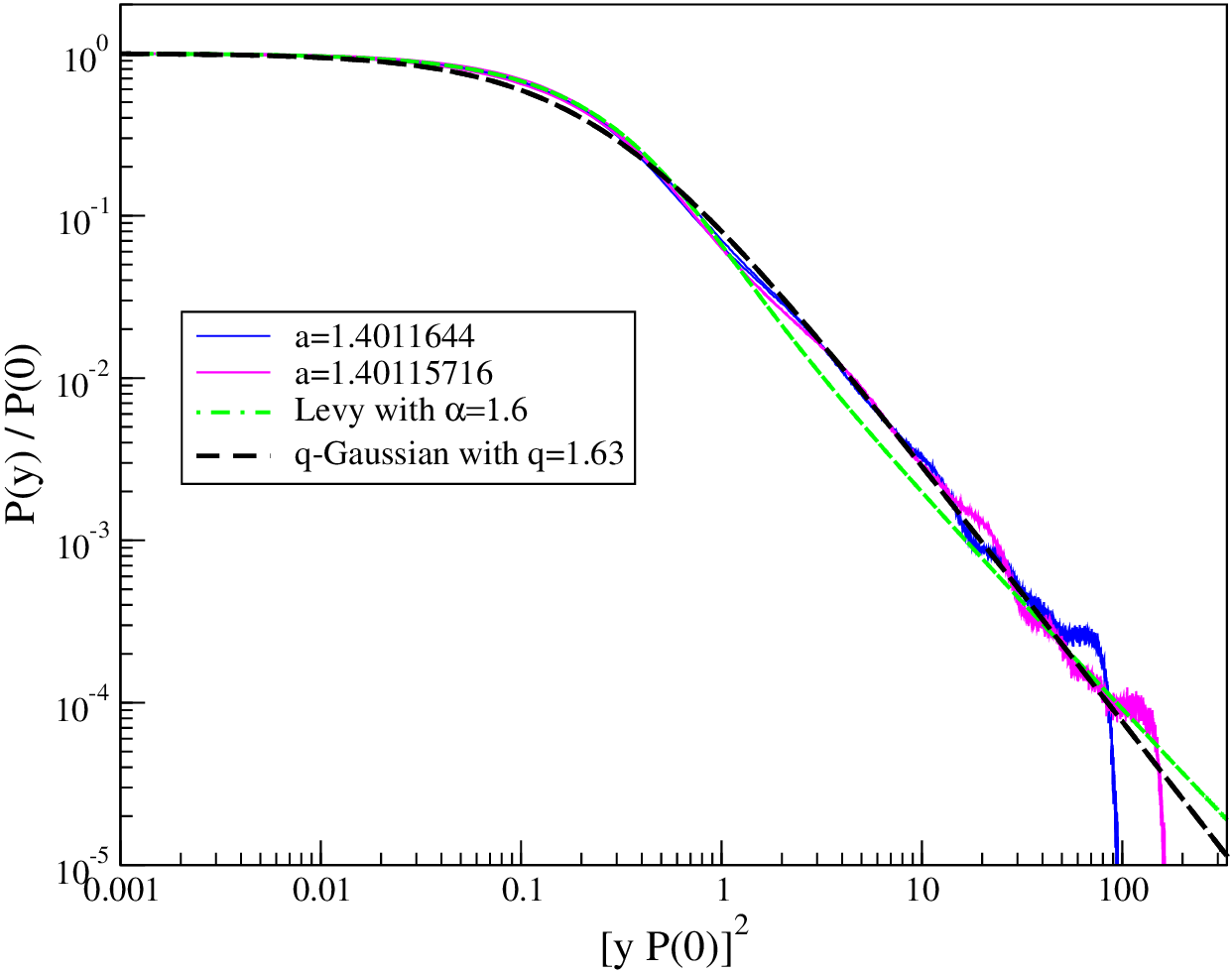}
\caption{Comparison of best fits obtained by
using $q$-Gaussians ($q=1.63$)
and L\'evy distributions ($\alpha =1.60$), both
in log-linear (a) and log-log plots (b).}
\end{figure}
The numerical data
are well approximated by $q$-Gaussians, whereas  L\'evy distributions
give worse fits. More specifically, in a log-log plot, 
L\'evy distributions with $1<\alpha<2$ have an inflexion point 
which is by no means supported by the logistic-map data.
If the parameter $\alpha$ of the
L\'evy distribution is slightly increased, then
the fit quality in the
middle region is slightly improved but the tails become too pronounced
to provide an acceptable fit.
Hence the claim of \cite{grass} that L\'evy distributions
might yield a better fit is incorrect.
Besides this, the L\'evy-Gnedenko limit theorem holds for {\em independent} (or nearly so)
random variables with infinite variance, whereas the iterates of the
logistic map near to the critical point have strong correlations.
Hence there is no theoretical justification to use L\'evy distributions
in this problem.
The conjecture in \cite{grass} that there might be a suitable ordering
of the iterates into subsets that are almost independent
lacks any theoretical proof or numerical justification.

Finally, in Fig.~1 of \cite{grass}, new types of distributions
of $Y$ are shown for the case $N_0=0$, i.e. no
transients are omitted, and claims are made at the end of
the paper that these
distributions including all the transients could
be universal. As an argument for universality of transients
at $a=a_c$,
in \cite{grass}  the work
\cite{scheunert} is cited. In \cite{scheunert}, however,
only transient behaviour of iterates of the exact Feigenbaum
fixed point function $g$ is investigated, i.e. the
map under consideration in \cite{scheunert} is the exact solution $g$ of
the Feigenbaum-Cvitanovic equation $\alpha g(g(x/\alpha))= g(x)$.
However, universal behavior
in our case would mean that different maps $f$ with quadratic maximum
would generate the same distribution of $Y$. Since different quadratic maps
can have
very different
transient behavior, it seems highly unlikely that the sum $Y$ of
all these different transients would converge to a random variable
that has a universal distribution, as claimed in \cite{grass}.
For this, one would have to carefully estimate the speed of convergence
of the iterates of $f$ to the Feigenbaum fixed point function $g$
under successive iteration and rescalation,
which was not done in \cite{grass}.
Hence there is no theoretical basis for the claim
of \cite{grass} that the observed
{\em transient} distributions of sums are universal.
Neither any numerical evidence of universality is provided
in \cite{grass}.

Summarizing, 
the numerical experiments of \cite{grass} were
mainly performed in a different parameter region
that was explicitly excluded by our scaling relation derived in \cite{tibet2}.
In the parameter region chosen by Grassberger the number $N$
of iterations  
is insufficient. 
In the region fixed by our scaling argument $N\sim n^2$,
$q$-Gaussians indeed provide good fits 
of the data, far better than the L\'evy distributions suggested
in \cite{grass}, which moreover do not have any theoretical justification
for this problem involving strongly correlated random variables.
Transient distributions investigated in \cite{grass} are unlikely to be universal.

This work has been supported by TUBITAK (Turkish Agency) under the Research Project number 104T148.
C.T. acknowledges partial financial support from CNPq and Faperj (Brazilian Agencies).

\end{document}